\pdfoutput=1
\documentclass[10pt, conference, letterpaper]{IEEEtran}

\usepackage[T1]{fontenc}
\usepackage[utf8]{inputenc}
\usepackage{verbatim} %
\usepackage{xspace}
\usepackage[textsize=tiny]{todonotes}
\usepackage[binary-units]{siunitx}
\usepackage[USenglish]{babel}
\usepackage{amsmath}
\usepackage{amssymb}
\usepackage{mathtools}
\makeatletter
\let\MYcaption\@makecaption
\makeatother
\usepackage[font=footnotesize]{caption}
\usepackage[font=footnotesize]{subcaption}
\usepackage[inline]{enumitem}
\usepackage{float}
\usepackage{hyperref}
\usepackage{threeparttable}

\usepackage[draft=false,tracking=true]{microtype} %
\usepackage{booktabs}		%

\usepackage{setspace}

\makeatletter
\let\MYcaption\@makecaption
\makeatother

\usepackage{nth}			%
\usepackage{csquotes}		%

\usepackage[nolist]{acronym}

\usepackage[capitalise,nameinlink]{cleveref}
\crefname{section}{Sec.}{Sec.}
\Crefname{section}{Section}{Sections}
\crefname{equation}{eq.}{eq.}
\crefname{figure}{Fig.}{Fig.s}
\Crefname{figure}{Figure}{Figures}

\usepackage[%
 backend=biber,
 style=ieee,
 url=false,
 isbn=false,
 doi=false,
 maxcitenames=1,
 mincitenames=1,
 maxbibnames=5,
]{biblatex}
\renewcommand*{\bibfont}{\footnotesize}

\AtEveryBibitem{\clearfield{note}}
\AtEveryBibitem{\clearlist{location}}
\AtEveryBibitem{\clearfield{pages}}
\AtEveryBibitem{\clearfield{series}}
\AtEveryBibitem{\clearfield{month}}

\newcommand*{\eg}{e.\,g.\@\xspace}
\newcommand*{\ie}{i.\,e.\@\xspace}
\newcommand{\R}{\mathbb{R}}
\newcommand{\N}{\mathbb{N}}
\newcommand{\Pb}{\mathbb{P}}

\newcommand{\dd}{\mathrm{d}}
\newcommand{\bs}{\textit{BitSwap}}

\newcommand{\wl}{\texttt{want\_list}}
\newcommand{\mtwb}{\texttt{WANT\_BLOCK}}
\newcommand{\mtwh}{\texttt{WANT\_HAVE}}
\newcommand{\mtwbsdh}{\texttt{WANT\_BLOCK\_SEND\_DH}}
\newcommand{\mtwhsdh}{\texttt{WANT\_HAVE\_SEND\_DH}}
\newcommand{\mtcancel}{\texttt{CANCEL}}

\newcommand{\mthave}{\texttt{HAVE}}
\newcommand{\mtdonthave}{\texttt{DONT\_HAVE}}
\newcommand{\mtblock}{\texttt{BLOCK}}

\newcommand{\mdag}{Merkle DAG}
\newcommand{\codec}{Multicodec}

\newcommand{\node}{v}
\newcommand{\dataitem}{d}
\newcommand{\cid}{c}
\newcommand{\session}{S(\cid)}
\DeclareMathOperator{\addr}{addr}
\newcommand{\providers}{P(\cid)}
\DeclarePairedDelimiter\len{\lvert}{\rvert}

\newcommand{\detailedweek}{from April \nth{30} to May \nth{06} 2021}
\newcommand{\detailedweekwofrom}{April \nth{30} to May \nth{06} 2021}

\newcommand{\inlinecode}[1]{\texttt{#1}}

\usepackage{tikz}
\usetikzlibrary{positioning, graphs, shapes.multipart, fit, calc, shapes.misc, matrix, shapes}
\usetikzlibrary{arrows, automata, plotmarks}

\definecolor{hublue} {RGB}{  0, 55,108}
\definecolor{hured}  {RGB}{138, 15, 20}
\definecolor{hugreen}{RGB}{  0, 87, 44}
\definecolor{husand} {RGB}{210,192,103}
\definecolor{hugraygreen}{RGB}{209,209,194}
\definecolor{hugrayblue} {RGB}{189,202,211}

\definecolor{jwigreige} {RGB} {166, 157, 130}
\definecolor{jwimauve} {RGB} {125, 80, 90}
\definecolor{jwiblue} {RGB} {35, 90, 130}
\definecolor{jwidarkgreen} {RGB} {70, 105, 90}
\definecolor{jwilightgreen} {RGB} {130, 150, 100}
\definecolor{jwiyellow} {RGB} {200, 140, 40}
\definecolor{jwiorange} {RGB} {190, 85, 45}

\begin{document}

\begin{acronym}[Derp]
\acro{p2p}[P2P]{peer-to-peer}
\acro{dht}[DHT]{distributed hash table}
\acro{ipfs}[IPFS]{Interplanetary Filesystem}
\acro{ipld}[IPLD]{Interplanetary Linked Data}
\acro{json}[JSON]{JavaScript Object Notation}
\acro{sfs}[SFS]{Self-Certifying Filesystem}
\acro{ipns}[IPNS]{Interplanetary Namesystem}
\acro{cid}[CID]{content identifier}
\acro{dag}[DAG]{directed acyclic graph}
\acro{protobuf}[Protobuf]{Protocl Buffers}
\acro{cdf}[CDF]{cumulative distribution function}
\acro{pdf}[PDF]{probability density function}
\acro{nft}[NFT]{non-fungible token}
\end{acronym}

\title{Monitoring Data Requests in Decentralized Data Storage Systems: A Case Study of IPFS}

\author{\IEEEauthorblockN{Leonhard Balduf\IEEEauthorrefmark{1}\IEEEauthorrefmark{2}\IEEEauthorrefmark{3},
		Sebastian Henningsen\IEEEauthorrefmark{1}\IEEEauthorrefmark{3},
		Martin Florian\IEEEauthorrefmark{1}\IEEEauthorrefmark{3},
		Sebastian Rust\IEEEauthorrefmark{2},
                Björn Scheuermann\IEEEauthorrefmark{1}\IEEEauthorrefmark{2}}
	\IEEEauthorblockA{\IEEEauthorrefmark{1}Weizenbaum Institute for the Networked Society, Berlin, Germany}
	\IEEEauthorblockA{\IEEEauthorrefmark{2}Technical University of Darmstadt, Darmstadt, Germany}
	\IEEEauthorblockA{\IEEEauthorrefmark{3}Humboldt University of Berlin, Berlin, Germany}}

\maketitle

\begin{abstract}
  Decentralized data storage systems like the \ac{ipfs} are becoming increasingly popular, \eg, as a data layer in blockchain applications and for sharing content in a censorship-resistant manner.
  In \ac{ipfs}, data is hosted by an open set of nodes and data requests are broadcast to connected peers in addition to being routed via a \ac{dht}.
  In this paper,
  we present a passive monitoring methodology that
  exploits this design for obtaining
  data requests from a significant and upscalable portion of nodes.
  Using an implementation of our approach for the \ac{ipfs} network
  and data collected over a period of fifteen months,
  we demonstrate how our methodology enables profound insights into,
  among other things:
  the size of the \ac{ipfs} network,
  activity levels and structure, and
  content popularity distributions.
  We furthermore present that our methodology can be abused for attacks on users' privacy.
  For example, we were able to identify and successfully surveil the \ac{ipfs} nodes corresponding to public \ac{ipfs}/HTTP gateways.
  We give a detailed analysis of the mechanics and reasons behind implied privacy threats and discuss possible countermeasures.

\end{abstract}

\section{Introduction}
\label{sec:intro}

Decentralized, peer-to-peer-based data storage systems are becoming increasingly popular, especially in the context of blockchain applications and censorship-resistant data hosting.
Whereas the narratives around previously conceived systems such as BitTorrent were mainly focused on performance and scalability, newer projects put a stronger emphasis on resilience and censorship resistance.
This shift in narratives is also reflected in the systems' designs, \eg, in an increased usage of techniques common to unstructured overlay networks.
The \acf{ipfs} is a prominent example of a newer \ac{p2p} data storage system~\cite{benet2014ipfs}.
\ac{ipfs} is in active real-world use for mirroring censorship-threatened content
and forms the data storage layer of various blockchain-based applications, including \ac{nft} platforms%
\footnote{For example: \url{https://opensea.io/}}.
As laid out by previous works~\cite{henningsen2020mapping,henningsen2020crawling}, \ac{ipfs} employs a hybrid approach between a structured Kademlia overlay and broadcasting of so-called \bs{} requests for content to directly connected peers.
While this combination renders \ac{ipfs} reasonably robust against, \eg, eclipse attacks~\cite{singhDefendingEclipseAttacks2004}, it enables extensive monitoring, putting individual users' privacy at risk.

In this paper, with \ac{ipfs} as a case study,
we explore unintended consequences of these robustness-increasing design features. %
We present (1) a passive monitoring methodology for collecting and processing \bs{} data requests of a large share of the network and (2) a monitoring setup as an instance of the methodology.
Our system enables us to reveal \emph{who} requested \emph{which} data item \emph{when}, \ie, which nodeID requested which \ac{cid} at what timestamp.
We collected measurements for fifteen months using two spatially diverse monitoring nodes,
yielding traces of %
$2.78 \cdot 10^{10}$
 data request entries in total.
In this work, we use the obtained dataset to highlight possible analysis angles.
Based on excerpts of our dataset, we showcase:
\begin{itemize}
  \item estimations of the size of the network
    that also account for non-\ac{dht} nodes
    (unlike previous methods),
  \item analyses of activity levels and structure (\eg, geography-based usage patterns),
  \item the derivation of content popularity distributions, and
  \item the feasibility of privacy attacks, by identifying node IDs of HTTP/\ac{ipfs} gateways.
\end{itemize}

An adversary with a similar setup to ours can determine (1) which nodes are interested in a given \ac{cid}, (2) which \acp{cid} were requested by a particular node, and, with negligible deniability, (3) whether a node (and hence its user) has downloaded a specific (\ac{cid}-referenced) data item in the recent past.
We follow up by a discussion of the design space of countermeasures to these privacy threats, highlighting promising directions for further investigation.

In summary, our main contributions are threefold:
\begin{itemize}
  \item a methodology for monitoring and processing content-related activity data (\cref{sec:monitoring}),
  \item results from a measurement study based on this methodology that showcase the utility of collectable data (\cref{sec:results}),
  \item a discussion of privacy risks implied by our methods and potential privacy-enhancement approaches (\cref{sec:privacy_risks}).
\end{itemize}

Our work highlights privacy-related issues in the popular \ac{p2p} system \ac{ipfs}.
While it is no secret that \ac{ipfs} is not a privacy-focused network\footnote{See, for example: \url{https://docs.ipfs.io/concepts/privacy-and-encryption/}},
our work demonstrates that the privacy level it offers is in some ways worse than that of the standard, \enquote{non-distributed} web. %
To safeguard and inform users, and in consultation with an ethics committee at HU Berlin,
we have publicly committed to a privacy policy\footnote{
  \url{https://monitoring.ipfs.trudi.group/privacy_policy.html}
} that describes our data collection and processing practices
and limits them to the purpose of deriving general, non-personalized insights and statistics about the IPFS network.
\section{Related Work}
\label{sec:related_work}

A variety of \ac{p2p} systems have been measured and monitored in the past,
through passive measurements, active crawling and probing, and combined approaches.
Classic studies include works on the Gnutella network~\cite{saroiu2001measurement,stutzbachCapturingAccurateSnapshots2006} that also highlight the pitfalls of churn when measuring dynamic networks.
BitTorrent~\cite{DBLP:conf/p2p/WangK13,pouwelseBittorrentP2PFileSharing2005} and the KAD \ac{dht}~\cite{steinerGlobalViewKad2007,steinerLongTermStudy2009} have similarly been subjects of investigation.
\textcite{DBLP:conf/p2p/WangK13} highlight the importance of using more than one vantage point and discuss important considerations when combining network size estimates from crawls.

More recently, \ac{p2p} networks have received renewed attention in the context of cryptocurrencies such as Bitcoin and Ethereum~\cite{tschorsch2016bitcoin}.
In a measurement setup similar to ours, \textcite{neudecker2016timing} inferred the topology of the Bitcoin overlay through monitoring block and transaction distributions from several spatially diverse nodes.
A similar approach is repeated in~\cite{ben2018vivisecting}.
Network-level measurements for topology inference have also been conducted for the privacy-focused cryptocurrencies Monero~\cite{DBLP:conf/fc/CaoYDLV20} and ZCash~\cite{DBLP:conf/lcn/DanielRT19}.
Apart from topology inference, the latency and bandwidth of Bitcoin and Ethereum peers were measured to assess the systems' degree of centralization~\cite{DBLP:conf/fc/GencerBERS18} and to assess the network health of Ethereum in general~\cite{DBLP:conf/imc/KimMMMMB18,DBLP:conf/iscc/0003SWTZY19}.

\ac{ipfs}, although significantly larger (in terms of node numbers) than most cryptocurrency networks~\cite{henningsen2020mapping}, has received comparatively little attention so far.
Previous studies have focused on the performance of data delivery~\cite{ascigilPeertoPeerContentRetrieval2019,UnderstandingPerformanceIPFS,confaisObjectStoreService2017}, the performance of DTube,
an application on top of \ac{ipfs}~\cite{DBLP:conf/networking/DoanPOB20},
and the susceptibility of \ac{ipfs} to eclipse attacks~\cite{DBLP:journals/corr/abs-2011-00874}.

In our own previous works~\cite{henningsen2020mapping,henningsen2020crawling},
we investigated \ac{ipfs}' overlay structure by crawling the Kademlia-based~\cite{maymounkov2002kademlia} \ac{dht}.
We found that clients maintain an unexpectedly large number of connections, a subset of which is stored in the $k$-buckets of the underlying \ac{dht} and can be assessed through crawling.
Instead of relying only on standard \ac{dht} searches,
nodes also broadcast data request to \emph{all} of their immediate overlay neighbors.
This insight forms the basis of our following investigation, as we exploit \ac{ipfs}' broadcasting behavior through passively collecting request messages from peers.
\section{IPFS in a Nutshell}
\label{sec:ipfs_in_a_nutshell}

In the following,
we give a concise overview of the popular decentralized data storage system \ac{ipfs},
based on our study of its source code and the available documentation.
We focus on aspects relevant to our monitoring methodology and subsequent observations.
Note that the development of \ac{ipfs} and related libraries is ongoing---we focus on inherent conceptual properties here as details of the design may change over time.
As a very broad overview, the design of \ac{ipfs} can be summarized in the following way:

\begin{itemize}
  \item \ac{ipfs} is a permissionless system with weak identities;
    anyone can deploy a node on the \ac{ipfs} overlay network.
  \item \ac{ipfs} nodes are identified by the hash of their public key, $H(k_{\text{pub}})$.
  \item Each data item on \ac{ipfs} is stored and served by one or more \emph{data providers} (nodes in the \ac{ipfs} overlay).
  \item References to data providers are stored in a Kademlia-based \ac{dht}
    (see~\cite{henningsen2020mapping,henningsen2020crawling} for more details on \ac{ipfs}' \ac{dht} and measurements thereof).
  \item Nodes are either \ac{dht} servers or \ac{dht} clients.
    The latter \emph{use} the \ac{dht} but are not part of the \ac{dht} network.
  \item Data items are transferred from providers using \ac{ipfs}' \emph{BitSwap} subprotocol,
    which functions similarly to BitTorrent~\cite{pouwelseBittorrentP2PFileSharing2005} and
    Bitcoin's inventory mechanism~\cite{nakamoto2008bitcoin,tschorsch2016bitcoin}.
  \item Data items are requested from \emph{all} connected overlay neighbors and the \ac{dht} is queried for providers only \emph{after} no neighbors were able to offer the data.
  \item By default, nodes \emph{cache} data items they have downloaded and effectively become a data provider for them.
\end{itemize}

In the following, we give a deeper introduction into
node types, %
data addressing, data retrieval,
and the \bs{} subprotocol.

\subsection{DHT Servers and DHT Clients}%
\label{sub:dht_servers_and_dht_clients}

Release \inlinecode{v0.5} of the \ac{ipfs} software introduced the separation of nodes into two distinct types: \ac{dht} servers and \ac{dht} clients.
The \ac{ipfs} software decides which mode to operate in based on whether it finds itself connectable from the Internet.
\ac{dht} servers are regular \ac{dht} participants: They store data that is inserted into the \ac{dht} and respond to \ac{dht} requests.
\ac{dht} clients
only use the \ac{dht}:
They do not store \ac{dht} data or process requests, and will not be included in $k$-buckets of other nodes.
Because of this, \ac{dht} clients cannot be enumerated using \ac{dht} crawling such as in~\cite{henningsen2020mapping}.

\subsection{Content Identifiers (CIDs) and Data Integrity} %
\label{sub:sfs}

\ac{ipfs} uses a form of \ac{sfs} \cite{mazieres00selfcert} to ensure the integrity of data throughout its delivery.
To this end, each data item $\dataitem$ is assigned a unique immutable address that is the hash of its content, \ie, $\addr(\dataitem) = H(\dataitem)$.
Recipients can recognize whether received data was tampered with by comparing its hash with the requested address.
In \ac{ipfs}, an $\addr(\dataitem)$ is encoded as a so-called \emph{content identifier} (CID).

Directories and files are organized as a \mdag{}\footnote{
  \url{https://docs.ipfs.io/concepts/merkle-dag/}
}.
This construction differs from Merkle Trees insofar as nodes can have more than one parent and, in the case of \ac{ipfs}, data on non-leaf nodes is permitted.
For example, a directory on \ac{ipfs} is encoded as a node containing the hashes of all entries in the directory in addition to metadata about each entry.
Large files are chunked into smaller data \emph{blocks} and encoded as multi-layered \ac{dag}s.
This construction ultimately allows for caching and deduplication of both file contents and directory entries.

In principle, \ac{ipfs} can be used to store a variety of different contents.
The encoding of a data item can be derived from its \ac{cid}, using a mapping known as \codec{}.
Important \codec{}s for \ac{ipfs} are:
\begin{enumerate*}
        \item \emph{DagProtobuf}, which encodes nodes for the \ac{ipfs} \mdag{}.
	These objects usually encode files and directories on \ac{ipfs}.
        \item \emph{Raw}, which are unencoded chunks of binary data or leaves of file Merkle \acp{dag}.
        \item \emph{DagCBOR} and \emph{DagJSON}, which are to-be replacements for DagProtobuf.
	They encode a generalized data model for hash-linked data, called \ac{ipld}, in different formats.
\end{enumerate*}
\subsection{Content Retrieval}
\label{sub:content_retrieval}

Data items are usually stored at multiple nodes in the network.
Nodes store content because they are its original authors, because they chose to \emph{pin} it, or because they have recently retrieved it themselves.
Nodes normally serve the data items they store upon request.
The nodes that store a given data item are consequently referred to as that data item's \emph{providers}.

\begin{figure}
  \centering
  \scalebox{0.55}{%
    \begin{tikzpicture}[node distance=1cm,
	arrow/.style={thick,->,>=stealth},
	decision/.style={diamond, minimum width=3cm, minimum height=1cm, aspect=2, text centered, draw=black, fill=green!30},
	process/.style={rectangle, minimum width=3cm, minimum height=1cm, text centered, draw=black, fill=orange!30},
	io/.style={trapezium, trapezium left angle=70, trapezium right angle=110, minimum width=3cm, minimum height=1cm, text centered, draw=black, fill=blue!30},
	startstop/.style={rectangle, rounded corners, minimum width=3cm, minimum height=1cm,text centered, draw=black, fill=red!30},
	every text node part/.style={align=center}]

		\node (req) [startstop] {User requests CID $c$.};

		\node (localsearch) [process, below=of req] {Search for $c$ in the local cache.};
		\draw [arrow] (req) -- (localsearch);
		
		\node (localsearchresult) [decision,below=of localsearch] {Found?};
		\draw [arrow] (localsearch) -- (localsearchresult);
		
		\node (locallyfound) [startstop, right=of localsearchresult] {Return data to user.};
		\draw [arrow] (localsearchresult) -- node[anchor=south] {yes} (locallyfound);
		
                \node (bcast-wh) [process, below=of localsearchresult] {Create a session $S(c)$ for $c$.\\
			Broadcast \mtwh{} $c$\\
			to \emph{all} connected peers.
                };
		\draw [arrow] (localsearchresult) -- node[anchor=west] {no} (bcast-wh);
		
		\node (recv-have) [decision, below=of bcast-wh] {Received \mthave{}s?};
		\draw [arrow] (bcast-wh) -- (recv-have);
		
                \node (addtosession) [process, below=of recv-have] {Add \mthave{}-sending peers to $S(c)$.};
		\draw [arrow] (recv-have) -- node[anchor=west] {yes} (addtosession);
		
		\node (search-providers) [process, above right=of recv-have,xshift=2cm,yshift=1cm] {Search the DHT\\
                        for providers $P(c)$.};
		\draw [arrow] (recv-have) -| node[anchor=south east] {no} (search-providers);
		
		\node (providers-dec) [decision, right=of search-providers] {Providers found?};
		\draw [arrow] (search-providers) --(providers-dec);
		
                \node (providers-found) [process,below=of providers-dec] {Establish connections to all $p \in P(c)$.\\%
			Broadcast \mtwh{} $c$\\
			to newly connected peers.%
		};
		\draw [arrow] (providers-dec) -- node[anchor=west] {yes} (providers-found);
	
		\node (providers-have) [decision, below=of providers-found] {Received \mthave{}s?};
		\draw [arrow] (providers-found) -- (providers-have);
		\draw [arrow] (providers-have) -| ++(-3cm,1cm) -- node[anchor=north] {yes} (addtosession.east);
		
		\node (idle-loop) [startstop, below=of providers-have] {Enter idle looping state.};
		\draw [arrow] (providers-have) -- node[anchor=west] {no} (idle-loop);
		
		\node (idle-loop-2) [startstop, above=of providers-dec] {Enter idle looping state.};
		\draw [arrow] (providers-dec) -- node[anchor=west] {no} (idle-loop-2);
		
		\node (send-want-block) [process, below=of addtosession] {Send \mtwb{} for $c$\\
                        to (some) peers in $S(c)$.
		};
		\draw [arrow] (addtosession) -- (send-want-block);
		
                \node (downloading) [right=of send-want-block] {\emph{(BitSwap until finished)}};
		\draw [arrow] (send-want-block) -- (downloading);
	
		\node (downloaded) [startstop, below=of downloading] {Return data to user.};
		\draw [arrow] (downloading) -- (downloaded);
		\end{tikzpicture} 
  }
  \caption{Content retrieval in \ac{ipfs}: broadcast and then query the \ac{dht}. The \enquote{idle looping state} entails periodically re-broadcasting a request for $c$ and re-searching the DHT.}
  \label{fig:content_retrieval_block_diagram}
\end{figure}
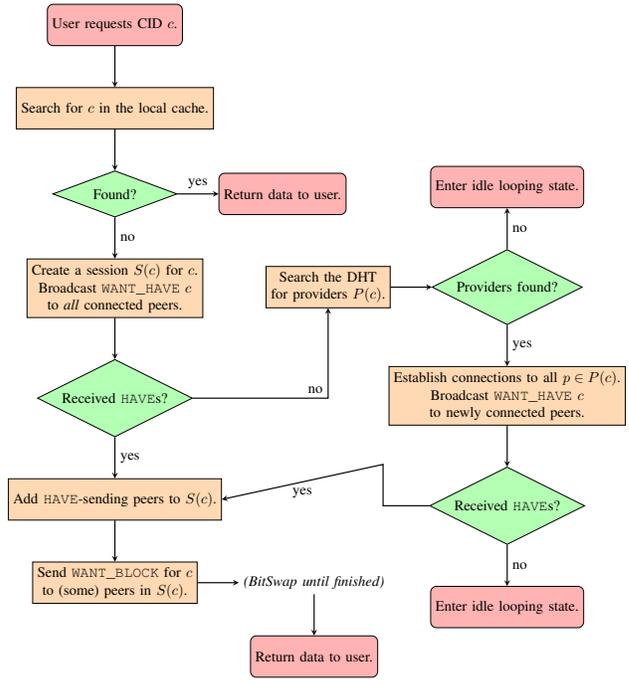

When an \ac{ipfs} node $\node$ wishes to retrieve a data item with CID $\cid$ (\eg based on a user request), it follows a two-step strategy (cf. \cref{fig:content_retrieval_block_diagram}):
\begin{enumerate}
  \item Ask \emph{all} nodes it is currently connected to for $\cid$, using the \bs{} subprotocol (\cref{sub:bitswap_explanation})
  \item If the first step fails, look up the providers $\providers$ for $\cid$ in the \ac{dht}, then request $\cid$ from members of $\providers$, again via \bs{}.
\end{enumerate}
Peers discovered through either stage are added to a \emph{session} $\session$, which is used to scope subsequent request for data related to $\cid$.
In the general case, $c$ initially references the root of a \ac{dag} of blocks,
which $v$ subsequently requests from the peers in the session.

One of the keystones of \ac{ipfs}' design is the caching and reproviding of requested blocks.
By default, the \ac{ipfs} node software stores up to \SI{10}{\giga\byte} of block data, %
with an optional garbage collection mechanism.
Users can also \emph{pin} \acp{cid} to ensure their local availability.
In this case, the given \ac{cid} and the \ac{dag} referenced by it is downloaded and marked exempt from garbage collection.

\subsection{BitSwap}
\label{sub:bitswap_explanation}

The \bs{} protocol is the main \enquote{data trading module}\footnote{
  \url{https://github.com/ipfs/go-bitswap}
} of \ac{ipfs}.
It is similar to BitTorrent’s~\cite{pouwelseBittorrentP2PFileSharing2005} inventory mechanism~\cite{tschorsch2016bitcoin} and is used for obtaining data items from connected peers.
\bs{} encompasses both
(1) announcing interest in \acp{cid} and discovering providers, and
(2) actually requesting and receiving the referenced data.
For both purposes, \bs{} builds upon a reliable transport layer such as TCP, QUIC, or even WebSockets.

\subsubsection{Inventory Mechanism}
\label{sec:ipfs:bitswap:inventory}

Upon user request to download a data item $\dataitem$, \ac{ipfs} broadcasts a \bs{} message to each connected peer.
Since \ac{ipfs} version \inlinecode{v0.5}, this message contains a \mtwh{} entry for a \ac{cid} $\cid = \addr(\dataitem)$, which can be understood as \enquote{I am looking for this block, do you have it?}.
Nodes receiving a request of this type answer with \mthave{} or \mtdonthave{}, depending on whether they have the content or not.
The latter is optional, a timeout mechanism alternatively determines absence of data.
Prior to \inlinecode{v0.5}, no inventory mechanism was present---data was requested directly, potentially leading to a high redundancy at the receiving side~\cite{ascigilPeertoPeerContentRetrieval2019}.
The interests of connected peers is tracked in so-called \wl{}s, which are persisted for as long as the peer is connected.
An outstanding request can be canceled with a \mtcancel{} entry, which is done on receipt of the requested data or on user request.

\subsubsection{Sessions}
\label{sec:ipfs:bitswap:sessions}

As shown in \cref{fig:content_retrieval_block_diagram}, recent versions of \bs{} operate on sessions.
A session $\session$ tracks the set of peers likely to have data related to a running query, based on receipt of \mthave{} messages and \ac{dht} searches.
Future requests for blocks related to $\dataitem$ can be directed at the relevant peers rather than flooded to all connected peers.
If no progress on a download is made, \ac{ipfs} attempts to extend the session with repeated broadcasts and \ac{dht} searches.

\subsubsection{Transmitting Data}
\label{sec:ipfs:bitswap:transmitting_data}

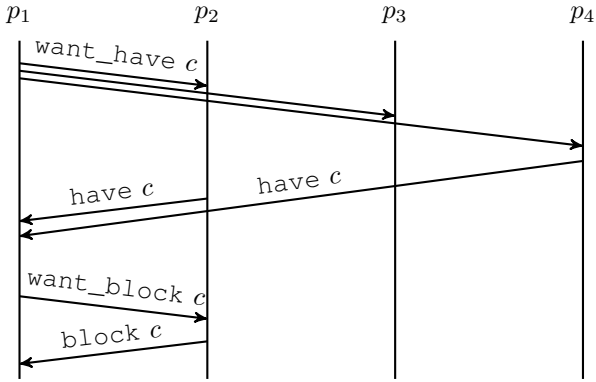
\begin{figure}
	\centering
	\begin{tikzpicture}
[>=stealth',thick,
commentl/.style={text width=3cm, align=right},
commentr/.style={commentl, align=left},]

\node (init) {$p_1$};
\node (p2) [right=1.9cm of init] {$p_2$};
\node (p3) [right=1.9cm of p2] {$p_3$};
\node (p4) [right=1.9cm of p3] {$p_4$};

\draw[->] ([yshift=-.4cm]init.south) coordinate (fn1o) -- ([yshift=-.3cm]fn1o-|p2) coordinate (fn1e) node[pos=.5, above, sloped] {\texttt{want\_have} $c$};
\draw[->] ([yshift=-.5cm]init.south) coordinate (fn2o) -- ([yshift=-0.7cm]fn1o-|p3) coordinate (fn2e) node[pos=.5, above, sloped] {};
\draw[->] ([yshift=-.6cm]init.south) coordinate (fn3o) -- ([yshift=-1.1cm]fn1o-|p4) coordinate (fn3e) node[pos=.5, above, sloped] {};

\draw[->] ([yshift=-1.5cm]fn1e) coordinate (resp1o) -- ([yshift=-.3cm]resp1o-|init) coordinate (resp1e) node[pos=.5, above, sloped] {\texttt{have} $c$};
\draw[->] ([yshift=-.2cm]fn3e) coordinate (resp3o) -- ([yshift=-.5cm]resp1o-|init) coordinate (resp3e) node[pos=.5, above, sloped] {\texttt{have} $c$};

\draw[->] ([yshift=-.8cm]resp3e) coordinate (fn2o) -- ([yshift=-.3cm]fn2o-|p2) coordinate (fn2e) node[pos=.5, above, sloped] {\texttt{want\_block} $c$};

\draw[->] ([yshift=-.3cm]fn2e) coordinate (resp2o) -- ([yshift=-.3cm]resp2o-|init) coordinate (resp2e) node[pos=.5, above, sloped] {\texttt{block} $c$};

\draw[thick, shorten >=-.2cm] ([yshift=-.1cm]init.south) -- (init|-resp2e);
\draw[thick, shorten >=-.2cm] ([yshift=-.1cm]p2.south) -- (p2|-resp2e);
\draw[thick, shorten >=-.2cm] ([yshift=-.1cm]p3.south) -- (p3|-resp2e);
\draw[thick, shorten >=-.2cm] ([yshift=-.1cm]p4.south) -- (p4|-resp2e);

\node (dummydots) [below right=.1cm and 1cm of resp2e] {       };

\node[right =2mm of fn3e, align=left] {%
};

\node[right =2mm of fn2e] {%
};
\end{tikzpicture}
        \caption{Obtaining a block with CID $\cid$ via BitSwap.}%
	\label{fig:bitswap_sequence_diagram}
\end{figure}

A node requests a block with a \mtwb{} entry for the block's \ac{cid}, which similarly can be canceled with a \mtcancel{} entry.
This request type is the backwards-compatible formalization of \enquote{I am looking for this block, send it to me if you have it}, which has been present in all versions of \bs{}.
If the peer has the block, it responds with \mtblock{} containing the data.
There is no negative response, a timeout tracks absence of data.
\Cref{fig:bitswap_sequence_diagram} depicts a typical procedure for obtaining a data block with CID $\cid$ using both request types
(node $p_1$ resolves $\cid$ via its peers $p_2, p_3, p_4$).

\section{Monitoring Data Requests}%
\label{sec:monitoring}

Decentralized data storage systems like \ac{ipfs} are inherently hard to monitor.
Signaling messages and data are exchanged directly between peers,
without passing through centrally-controlled infrastructure that could form a natural vantage point.
In previous works, we demonstrated how, using crawling, the nodes and connections forming the \ac{ipfs} \ac{dht}
can nevertheless be made visible~\cite{henningsen2020mapping, henningsen2020crawling}.
In the following, we present a methodology for monitoring \emph{data-related activity}---how many and which nodes request which data.
Notably, this also enables the systematic investigation of stored content.
In \ac{ipfs},
providers only return data when asked for the correct \ac{cid},
so in order to investigate stored content one must first learn about valid \acp{cid}---which can be done by observing data requests.

This section introduces our methodology for collecting, processing, and interpreting \bs{} data requests in \ac{ipfs}.
In \cref{sec:results} we apply our methodology
for conducting an exemplary measurement study that highlights the feasibility and potential of our approach.
While the detailed designs and empirical results we present in this paper are focused on \ac{ipfs},
our methods and conclusions are transferable to other decentralized data storage systems that share key design features,
most prominently the reliance on data request broadcasts.

\subsection{Data Collection}%
\label{sub:methodology}

Data collection can conceptually be described as a two-step process,
with each step leveraging different features of \ac{ipfs}' design.
Firstly, we \emph{operate nodes with infinite connection capacity}.
This is possible because, by design,
anyone can deploy a node on the \ac{ipfs} network and the number of connections a node can maintain is only limited by the \ac{ipfs} software.
Secondly, we \emph{collect all \bs{} messages} from connected nodes.
\ac{ipfs} nodes send data requests to all nodes they are connected to (cf. \cref{sec:ipfs_in_a_nutshell}),
and hence also to our \emph{monitoring nodes}.

By collecting the \bs{} traffic of a peer, we learn which \acp{cid} it requested, and at which times.
Our monitoring nodes produce, using a modified version of the \ac{ipfs} software,
a list of $(\texttt{timestamp}, \texttt{node\_ID}, \texttt{address}, \texttt{request\_type}, \texttt{CID})$ tuples.
No definite knowledge is gained about whether
(1) the data $d$ referenced by a \ac{cid} $c$ was downloaded successfully, and
(2) what $d$ \emph{is} (including whether it is a file or a directory).
The former can be determined by sending a request for $c$ to the requesting peer after it has issued a \mtcancel{} for $c$.
The latter can be determined by downloading and indexing $d$.

In the presented approach,
monitoring nodes are \emph{passive}.
They accept all incoming connections, but do not actively search for or connect to peers apart from usual node behavior (\eg, bootstrapping and \ac{dht} maintenance).
They thereby remain indistinguishable from regular nodes in terms of connection \emph{initiation} and generally do not send \bs{} requests or data.
In order to collect messages from a larger portion of the \ac{ipfs} network, multiple monitors with different node IDs can be used in conjunction.

Our collection methodology implies a number of limitations.
While enabling low-cost and hard-to-detect monitoring, passive monitors will generally only detect requests for root hashes of a \mdag{},
as requests further down in the \mdag{} are scoped to that \ac{cid}'s session (cf. \cref{sub:bitswap_explanation}).
Since our monitoring nodes $W$ do not hold any data, they are not added to any session and will therefore not receive any further requests.
\ac{ipfs} nodes furthermore cache downloaded data (cf. \cref{sub:content_retrieval}).
Subsequent request for the same data will be served from the local cache instead of being broadcast via \bs{}.
We can therefore observe only the first requests of a node for a given data item, or requests made after its cache was purged.
\subsection{Preprocessing}
\label{sec:measurement_setup:data_processing}

Each monitoring node produces a \emph{trace} of \bs{} messages it received.
If necessary,
traces from multiple monitors can be unified into one global trace.
If a node is connected to multiple monitors, we receive broadcast \wl{} entries multiple times.
To filter out these duplicates, we consider \wl{} entries received by different monitors to be identical if their source node ID,
request type, and target \ac{cid} match and their timestamps differ by at most \SI{5}{\second}.
The window size was chosen to account for most genuine duplicates, potentially delayed due to high latencies.

There are mechanisms in \ac{ipfs} that cause nodes to re-broadcast \wl{} entries every \SI{30}{\second} if the referenced data has not been downloaded yet
(cf. \cref{sec:ipfs:bitswap:sessions})%
\footnote{These messages serve little purpose, as \wl{}s are persisted---they seem to have historical origins.}.
These repeated broadcasts make up a significant portion of all requests ($>\SI{50}{\percent}$ according to our measurements),
skewing the numbers for some analyses.
We maintain another, larger, per-monitor window of \SI{31}{\second} to mark these duplicate messages.
Note that, as nodes maintain independent re-broadcast timers for each connected peer,
re-broadcast messages reach different monitors at shifted times.
This can lead to a misclassification of some same-monitor re-broadcasts as inter-monitor duplicates.

While in theory a balance between the \SI{5}{\second} and \SI{31}{\second} windows must be found,
in practice both are filtered out for the analyses presented here.

After data processing, we operate on a unified trace of $(\texttt{timestamp}, \texttt{node\_ID}, \texttt{address}, \texttt{request\_type}, \texttt{CID},\\ \texttt{flags})$ tuples, where the flags encode information about duplicate status and repeated broadcast detection\footnote{The corresponding tools and documentation are maintained at \url{https://github.com/mrd0ll4r/ipfs-tools}}.

\subsection{Estimating the Network's Size}%
\label{sub:methods:number_of_nodes}
The data we gather through our monitoring methodology can be used for estimating the total size $N$ of the \ac{ipfs} network.
For example, using two monitors \inlinecode{m1} and \inlinecode{m2},
and with the simplifying assumption that each monitoring node selects peer sets $P_\inlinecode{m1}$ and $P_\inlinecode{m2}$ uniformly and independently from the whole node population, we can estimate the size of the network as:
\begin{align}
	N_E = \frac{\len{P_\inlinecode{m1}} \cdot \len{P_\inlinecode{m2}}}{\len{P_\inlinecode{m1} \cap P_\inlinecode{m2}}},
  \label{eq:hypergeom_estimate}
\end{align}
where $N_E$ is an estimation for $N$.
This can be derived by considering the population of $N$ nodes as black balls in an urn, $K := \len{P_\inlinecode{m1}}$ of which are turned red through connections through \inlinecode{m1}.
Then, the connections of \inlinecode{m2} can be seen as sampling $n := \len{P_\inlinecode{m2}}$ balls without replacement---yielding a hypergeometric distribution with $k := \len{P_\inlinecode{m1} \cap P_\inlinecode{m2}}$ successes.
A maximum likelihood estimate of $N$ with the Stirling approximation ($\ln n! = n \ln n - n$) gives us $N_E = \frac{nK}{k} = \frac{\len{P_\inlinecode{m1}} \cdot \len{P_\inlinecode{m2}}}{\len{P_\inlinecode{m1} \cap P_\inlinecode{m2}}}$.

The general case of $r$ monitors can be handled through modeling the system as a coupon collectors problem with group drawings, also referred to as committee occupancy problem~\cite{mantel1968class}.
In this setting, peers correspond to cards and are numbered $1, \ldots, N$.
Assume each monitor has $w$ connections, then a monitor is a group drawing of $w$ cards/peers without replacement from the total set of peers---hence, there are no duplicates within a drawing but only between them.
We furthermore assume these drawings to be independent from one another.
Typically, the question modeled is ``What is the probability that we have $m$ distinct cards/peers after $r$ draws of size $w$?''.
Let $X$ be the number of distinct peers after $r$ draws of size $w$, then this probability is given by~\cite{mantel1968class}:
\begin{align}
  \Pb[X = m] = {N \choose w}^{-r} {N \choose m} \sum_{k=w}^m (-1)^{m - k} {m \choose k} {k \choose w}^r.
  \label{eq:collector_density}
\end{align}
In our setting, we know $m$ (the size of the union over all monitors' peer sets), and $N$ is the quantity to be estimated.
Hence, as for \cref{eq:hypergeom_estimate}, we can turn the probability density into a maximum likelihood estimation of $N$.
Given $m, r, w$, we (numerically) solve the following equation for $N$:
\begin{align}
  N - N \sqrt[r]{1 - \frac{m}{N}} - w = 0.
  \label{eq:mle_collector}
\end{align}

The accuracy of our estimation formulas is influenced by a number of factors.
Most prominently, due to various aspects of \ac{ipfs}' peer selection logic,
peer sets might not represent a uniform and independent draw from the whole node population.
The selection of peers is biased based on node IDs,
with node IDs close (in XOR metric) to the node ID of \inlinecode{m1} being more more likely to be connected to \inlinecode{m1} than nodes further away (\eg, \inlinecode{m2}).
Our measurements (presented in \cref{sub:number_of_nodes}) indicate that this is \emph{not} an important factor in our setup:
For long-running nodes with a high number of peers
the distribution of peers' node IDs is approximately uniform.

Our monitors' selection of peers
might also be biased based on more nuanced node characteristics.
Our monitors never evict peers and do not actively search for peers beyond \ac{ipfs}' periodic \ac{dht} refresh.
Consequently,
we observe that the majority of our monitors' connections are inbound and that their peers are more likely to be client nodes and popular gateway nodes,
\ie,
nodes that acquire new peers more frequently.
Therefore,
stable, long-living nodes that seldom initiate data requests will be underrepresented in peer draws,
which can lead to estimation errors. %

In theory,
our estimates can also be influenced by effects resulting from the combination of observations from multiple monitoring nodes.
For example, a node already connected to \inlinecode{m1} has one less slot for forming new connections and is therefore minimally less likely to be connected to \inlinecode{m2} as well.
As \ac{ipfs} nodes establish anywhere between 600 and 900 connections on average, the effect of occupied capacities is limited.
With respect to our $r > 2$ estimator, it must be pointed out that $w$ is not the same for all monitors, as they slightly differ in the number of connections.
Due to the small difference, explicitly modeling a heterogeneous $w$ would significantly increase the complexity of the model with questionable benefits.
Based on our empirical observations, neither of these two factors seem to have a significant impact in practice.

It is noteworthy to remember that 1) the \ac{ipfs} network consists of \ac{dht} servers and clients
(see \cref{sub:dht_servers_and_dht_clients}), and that 2) it has an unusual topology (see \cite{henningsen2020mapping}):
Although it is, in principle, constructed on top of a Kademlia \ac{dht}, participating nodes hold \emph{additional} connections not present in their $k$-buckets.
These are, for example, connections that were opened during a \ac{dht} search or download.
\ac{dht} clients are \emph{not} present in $k$-buckets and cannot be enumerated using traditional \ac{dht} crawling.
They do, however, appear as peers to our monitoring nodes due to these additional connections.
As such, \ac{dht} crawls can observe the core of the network, and our monitoring nodes can observe the entire network,
albeit as of now with some bias due to their passive
peer selection.
In \cref{sub:number_of_nodes} we apply our network size estimators in practice and compare the results with numbers obtained through \ac{dht} crawls.

\subsection{Content Popularity}%
\label{sub:methods:content_popularity}

Collected traces of \bs{} requests can be used to deduce the relative popularity of \acp{cid}, and hence the content they reference.
Knowledge of this popularity distribution is, \eg, an important building block for the formal analysis of cache hit ratios (especially relevant for \ac{ipfs} gateways)~\cite{fricker2012versatile}.
It furthermore allows for more realistic network simulations and user models. %
To this end, we define two different popularity scores,
one for capturing \ac{ipfs}' behavior \enquote{on the wire} and one for approximating user behavior.
For the former,
we define a \ac{cid}'s \emph{raw request popularity} (RRP) as the total number of requests received for a particular \ac{cid} over a given period.
This number is of interest to simulation studies and for improving the performance of \bs{}.
For approximating user behavior, we consider \emph{unique request popularity} (URP),
the number of distinct peers that requested a respective \ac{cid} in the given time period.
The motivation behind URP is that requests for a \ac{cid} coming from distinct peers indicate the corresponding data item's popularity among distinct users.

\section{Example Monitoring Study}
\label{sec:results}

We implemented our monitoring methodology and have been collecting data from the public \ac{ipfs} network since March 2020.
In the following, we describe our setup and present exemplary observations that it has allowed us to make,
showcasing the feasibility and utility of our approach.

\subsection{Monitoring Setup}
\label{sub:measurement_setup}

Our setup consists of a small number of monitoring nodes $W$, which function as outlined in \cref{sub:methodology}.
They are deployed on publicly reachable servers, \ie, not behind a NAT.
For this paper, $|W| = r = 2$\footnote{
	While a higher $r$ might result in a larger portion of the network's requests being recorded,
	we found that the intersection over union of \bs{}-active peers (whose messages we record)
	between our two nodes is, on average, $>70\%$.
}, with one node situated in Germany and one in the United States.
The nodes were running a modified version of the \ac{ipfs} client,
which was kept up-to-date within a few weeks of every new stable release of \ac{ipfs}.

\subsection{Data Collection}
\label{sub:results}

We collected \bs{} traces continuously for fifteen months,
yielding over \SI{3.5}{\tera\byte} of compressed traces.
We maintained only one monitoring node for the first two months of our observations.
The remaining time was surveyed with two nodes.
Since beginning the data collection, our monitors underwent minor configuration and version changes as well as some outages.
We classify these as minor (\SI{>5}{\second}) and major (\SI{>1}{\hour}) outages and count the number of days during which they occurred.
The \inlinecode{us} monitor in the U.S. was running for a total of %
$450$%
 days,
$17$%
 days out of which it experienced one or more major outages and an additional %
$15$%
 days out of which it experienced one or more minor outages.
The \inlinecode{de} monitor in Germany was running for a total of %
$385$%
 days, %
$18$%
 days out of which it experienced one or more major outages and an additional %
$5$%
 days out of which it experienced one ore more minor outages.
We observed more than %
$806$%
 million unique \acp{cid} over the entire fifteen months on one monitor.

The remainder of this section showcases analyses possible with our collected data,
and with data collectible using our monitoring methodology in general.
We present results for:

\begin{itemize}
  \item Estimating the size of the \ac{ipfs} network.
  \item Assessing the level and structure of data-related activity.
  \item Investigating the popularity of content stored on \ac{ipfs}.
\end{itemize}

We leave further analyses on the "file system" layer of \ac{ipfs} that are enabled through the learned \acp{cid},
\eg, of
the structure of \ac{ipfs}' data graph,
for future works.

While presenting analysis results,
we will mostly focus on an excerpt of our collected data that corresponds to the week \detailedweek.
Focusing our discussion on a single week allows us to highlight more fine-granular results and perform more complex analyses.
\subsection{Monitoring Coverage and Network Size}%
\label{sub:number_of_nodes}

In the following, we take a closer look at traces collected in the week \detailedweek{}.
Over this period, our two monitors saw %
$78011$%
 and %
$81423$%
 unique peers in total, respectively, for a union of %
$99147$%
 peers.
The monitors were connected to an average number of, respectively, %
$7132.56$%
 and %
$7798.82$%
 peers,
with the size of the union of unique peers being %
$9628.67$%
 on average.
Notably, averages and weekly totals differ significantly from each other,
which is in line with previous observations about churn in the \ac{ipfs} network~\cite{henningsen2020mapping}.
Only a small portion of the nodes we monitored in the above period were actively engaging in the \bs{} protocol, sending at least one request or cancel.
The two monitors saw,
respectively,
$6080$%
 and %
$6247$%
 unique \bs{}-active peers during the studied period, with a union of %
$7520$%
 unique \bs{}-active peers.
\begin{figure}
\centering
  \scalebox{0.8}{%
  \input{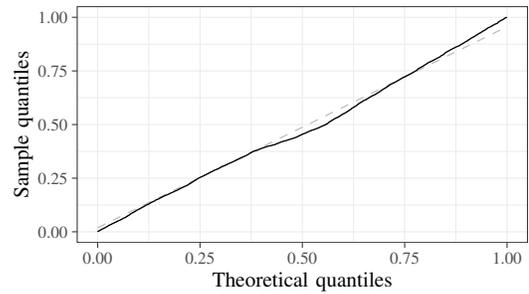}
  }
  \caption{Quantile-quantile plot of the IDs of peers connected to the \inlinecode{us} monitor in comparison to a uniform distribution (straight dashed line).
  }
  \label{fig:qqUniformity}
\end{figure}
These results allow us to estimate the network's size using the methods we propose in \cref{sub:methods:number_of_nodes}.
We first investigate whether the assumption of uniform peer selection
can reasonably be made.
We exemplarily gathered all connected peers of the \inlinecode{us} monitor on the \nth{4} of May\footnote{Other points in time yield similar results.}, totaling 8171 peers.
The comparison of the distribution of these node IDs to a standard uniform distribution is depicted in \cref{fig:qqUniformity}
as a quantile-quantile plot, with the straight dashed line corresponding to the uniform distribution.
It can be observed that the distribution of node IDs is surprisingly close to uniformity.
We then apply
our network size estimation formulas from \cref{sub:methods:number_of_nodes}.
Doing so
yields an estimated average network size of %
$10561$%
 (std. dev. %
$390$%
) with \cref{eq:hypergeom_estimate} and %
$10250$%
 (std. dev. %
$395$%
) with \cref{eq:mle_collector} (setting $w$ as the average of connections of both monitors).
While deriving a ground truth for this estimate is inherently challenging~\cite{DBLP:journals/jss/KostoulasPGBD07},
we can compare our results with alternative indicators for the \ac{ipfs} network's size.
Existing measurement infrastructure from previous works \cite{henningsen2020mapping,henningsen2020crawling} can be used to generate insights into the portion of the network reachable through crawls of the \ac{ipfs} \ac{dht}\footnote{
  See also: \url{https://trudi.weizenbaum-institut.de/ipfs_crawler.html}.
}.
Based on crawls of the \ac{ipfs} network during the discussed period, a total of %
$52463$%
 unique peers were found, with an average network size of %
$14411.42$%
 peers per crawl.
This hints at the fact that our method might in fact underestimate the current network size, as speculated in \cref{sub:methods:number_of_nodes}.
However, measuring the size of the \ac{ipfs} network based on \ac{dht} crawls has limitations on its own.
For example, crawled \ac{ipfs} nodes also propose \ac{dht} nodes to the crawler that are in fact offline~\cite{DBLP:journals/corr/abs-2011-00874}.
Such nodes are still counted by the crawler, as even online nodes might be unreachable if they reside behind NAT or other restrictive middleboxes.
On the other hand, our \ac{dht} crawler doesn't enumerate client-only nodes that are part of the \ac{ipfs} network but not part of the \ac{dht},
which potentially explains why our monitors saw more unique node IDs over the discussed week than the aforementioned crawler (%
$99147$%
 vs. %
$52463$%
).
As hinted at in our discussion in \cref{sub:methods:number_of_nodes}, the crawler and our monitors seem to be biased towards different parts of the \ac{ipfs} overlay, with monitors discovering more edge nodes
(that naturally generate more data request traffic)
and the \ac{dht} crawler a larger part of the core network.
Research opportunities remain for deepening the investigation of both approaches,
their biases and combination possibilities.
Still, the available estimations already enable a ballpark assessment of the number of nodes in the \ac{ipfs} network.
We can use the available network size estimates to gauge the \emph{coverage} of our monitoring approach.
We use the crawling-based estimation of the network's size in the following,
being the larger of the two and therefore more likely to underestimate our coverage.
At any given time, our monitoring nodes \inlinecode{us} and \inlinecode{de} were thereby connected to, and hence receiving \bs{} messages from,
\SI{54}{\percent} and %
\SI{49}{\percent} of the network, respectively.
The joint setup combining traces from both nodes had an average monitoring coverage of %
\SI{67}{\percent}.
Notably, we achieved this coverage using only two passive monitoring nodes.
The monitoring coverage can be further increased by adding more monitoring nodes or,
complementary, by implementing a more active peer discovery mechanism.

\subsection{Level and Structure of Data-Related Activity}%
\label{sub:general}

\begin{figure}
	\centering%
	\input{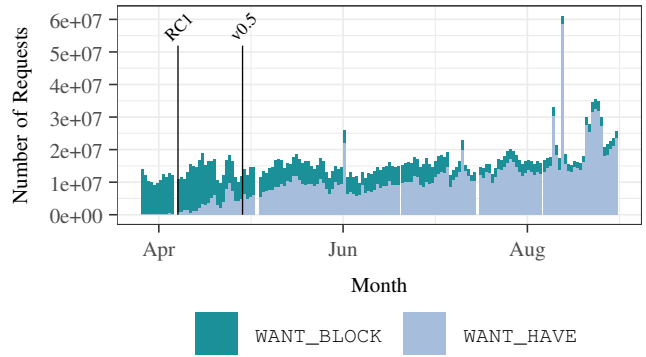}\unskip%
	\caption{
          Number of data requests collected in 2020 by monitor \inlinecode{us} over time,
          classified into the \mtwb{} and newer \mtwh{} request types.
          X-axis ticks mark the beginning of the indicated month. %
        }%
	\label{fig:message-types-over-time}
\end{figure}

Showcasing the potential of our methodology for monitoring the level and structure of data-related activity,
\cref{fig:message-types-over-time} depicts the view of monitor \inlinecode{us} on the number of requested \acp{cid} per day and entry type,
for the period from mid-March to the end of August 2020.
Requests are classified into the older \mtwb{} entry type and the \mtwh{} type officially introduced in \ac{ipfs} \inlinecode{v0.5}.
Missing values indicate incomplete data due to outages.
We observe a willingness of users to upgrade their clients. %
The large spike at the beginning of August was registered by both of our nodes, but we did not investigate further.

We also analyzed the collected \acp{cid} for the \codec{} they reference.
The \codec{} describes what type of data is referenced by a \ac{cid}, as outlined in \cref{sub:sfs}.
The results of our analysis over the entire fifteen months
are presented in \cref{tab:cid-codec-with-hash-count}.
Over this duration, we collected a total of %
$2.78 \cdot 10^{10}$
 data requests from both monitors.
We only count requested entries, not \mtcancel{}s, and derive the data from raw, unprocessed traces of the two monitors.
We can see that, from the perspective of our
monitoring infrastructure, \ac{ipfs} is used mostly for file storage.%

\begin{table}
  \begin{minipage}{0.24\textwidth}
        \centering
        \resizebox{\textwidth}{!}{
\begin{tabular}{lrr}
  \toprule
Codec & Count & Share (\%) \\ 
  \midrule
DagProtobuf & 23983048867 & $86.21$ \\ 
  Raw & 3734468407 & $13.42$ \\ 
  DagCBOR & 101649452 & $0.37$ \\ 
  GitRaw & 616657 & $<0.01$ \\ 
  EthereumTx & 163714 & $<0.01$ \\ 
   \multicolumn{1}{c}{Others (8)}&382982&$<0.01$\\
 \bottomrule
\end{tabular}

        }
	\caption[Number of \acp{cid} Received in Total by \codec{}]{
                Share of data requests by \codec{}. (March 2020--June 2021)
	}\label{tab:cid-codec-with-hash-count}
  \end{minipage}
  \hspace{0.01\textwidth}
  \begin{minipage}{0.22\textwidth}
      \vspace{-0.2mm}
        \centering
        \resizebox{0.55\textwidth}{!}{
	\begin{tabular}{lrr}
    \toprule
    Country & Share (\%) \\
    \midrule
    US  & $45.65$\\
    NL  & $13.85$\\
    DE  & $12.72$\\
    CA & $7.61$\\
    FR  &   $6.64$\\
    Others &  $<13.60$\\
    \bottomrule
\end{tabular}%

        }
	\caption[Share of Data Requests by Country.]{
                Share of data requests by country. %
                (April \nth{30}--May \nth{06} 2021)
	}\label{tab:geoipfs}
  \end{minipage}
  \vspace{-1.5em}
\end{table}

Our collected traces also allow insights into the geographic patterns of \ac{ipfs} usage.
We examined IP addresses from our unified, deduplicated dataset over the period between April 30th and May 06 2021.
and resolved them via the MaxMind GeoIP2 database\footnote{\url{https://dev.maxmind.com/geoip/geoip2/geolite2/}}.
\cref{tab:geoipfs} shows the share of observed \bs{} data requests per origin country.
We see that nodes residing in the US account for almost half of the observed activity and the top three countries together for roughly \SI{70}{\percent} of all observed activity during this period.

\subsection{Content Popularity}%
\label{sub:content_popularity}

Applying the popularity scores we defined in \cref{sub:methods:content_popularity} to the collected traces for the \detailedweekwofrom{} period
allows us to calculate the distribution of \ac{cid} popularities.
We compute both popularity metrics on the unified traces of both monitors, the resulting empirical CDFs (ECDF) are shown in \cref{fig:popularity_distributions}.
It can be seen that the majority of \acp{cid} in both distributions have a low popularity score, \eg, over \SI{80}{\percent} of \acp{cid} were only requested by one peer, as depicted in \cref{subsifg:urp_score}.
Although the distributions differ, both seem highly skewed with few highly-requested \acp{cid} and a majority of ``unpopular'' ones.
In contrast to other works on item popularity reporting heavy-tailed, Zipf-like distributions~\cite{fricker2012versatile}, fitting a power-law distribution to our measured scores (\cref{fig:popularity_distributions}) as laid out in~\cite{clauset2009power}
yields a $p$-value $< 0.1$, both for RRP and URP,
regardless of the choice for a cut-off value $x_{\min}$.
We therefore conclude that the power-law hypothesis has to be rejected, \ie, that the measured popularity data does likely \emph{not} follow a power-law distribution.

It has to be noted that popular data items according to RRP are often not resolvable, \ie, the data block the \acp{cid} are pointing to cannot be found.
This observation may stem from different factors.
First, \bs{} periodically re-broadcasts requests for \acp{cid} it cannot resolve (cf. \cref{sec:measurement_setup:data_processing}).
Furthermore, some peers issue an unexpectedly high number of requests for the same data item---hinting at configuration errors or non-standard usages of \ac{ipfs}.

The ten most popular \acp{cid} according to URP are resolvable.
These most popular \acp{cid} contain (1) various data related to the decentralized exchange Uniswap\footnote{\url{https://uniswap.org/}.} (\eg, their logo and config files), (2) JSON files related to running Ethereum nodes through the dAppNode-project\footnote{\url{https://github.com/dappnode/DAppNode}, a project devoted to simplifying the process of running full nodes in various \ac{p2p} networks.} and (3) a HTML page stating ``Unavailable for legal reasons''.
\begin{figure}
  \begin{subfigure}{0.25\textwidth}%
    \scalebox{0.5}{%
      \input{fig/rawEcdfs}
    }%
    \caption{RRP score}
    \label{subfig:rrp_score}
  \end{subfigure}%
  \begin{subfigure}{0.25\textwidth}%
    \scalebox{0.5}{%
      \input{fig/uniqEcdfs}
    }%
    \caption{URP score}
    \label{subsifg:urp_score}
  \end{subfigure}%
  \caption{ECDFs of content popularities (\detailedweek{}).}
  \label{fig:popularity_distributions}
\end{figure}

\section{Privacy Risks}%
\label{sec:privacy_risks}

The monitoring of data requests in \ac{ipfs}
provides useful insights for assessing the network's level of use, identifying key usage scenarios,
and tuning performance.
However, our monitoring techniques can also be used for tracking the behaviour of individual users,
implying latent privacy risks.
In the following, we flesh out these risks by proposing a series of specific attacks on the privacy of \ac{ipfs} users.
As a demonstration,
we uncover the (normally hidden) node identifiers corresponding to public \ac{ipfs}/HTTP gateways and successfully track requests initiated through these gateways.
We also discuss countermeasures,
highlighting that the identified attacks are in part enabled by core aspects of \ac{ipfs}'s design,
and that remedying them is a hard challenge when other desirables from a decentralized data storage system are taken into account.

\subsection{Privacy attacks on \ac{ipfs}}%
\label{sub:privacy_attacks_on_ipfs}

We define three specific attack vectors that can enable adversaries to learn about the current and past data request behaviour of \ac{ipfs} nodes:
\emph{Identifying Data Wanters} (IDW),
\emph{Tracking Node Wants} (TNW),
and
\emph{Testing for Past Interests} (TPI).
In the standard usage mode, \ac{ipfs} users access and distribute data via \ac{ipfs} nodes under their own control.
Consequently, learning what data a node is or has been interested in allows for direct conclusions about the (likely private) preferences of its human operator\footnote{
  Note that even if data items have been encrypted before being placed on \ac{ipfs},
  metadata such as request behaviours and approximate data sizes can still be learned by an adversary.
}.
We discuss the alternative, (public) gateway-based usage mode of \ac{ipfs} in \cref{sub:privacy_countermeasures}.

\subsubsection{Identifying Data Wanters (IDW)}%
\label{ssub:cid_wanters}

The goal of the IDW attack is to discover nodes that are interested in a given CID-identified data item.
The setup of the attack is identical to our monitoring setup. %
In fact, our deployed monitoring infrastructure
already collects the necessary information for listing node IDs that have requested a given CID.
By deploying more monitoring nodes or using an active, crawling-like peer discovery approach,
an adversary can increase the number of nodes to which he maintains a direct connection and from which he receives CID messages.
Already with one monitoring node, however,
we were able to monitor more than \SI{45}{\percent} of the public \ac{ipfs} network (cf. \cref{sub:number_of_nodes}).

\subsubsection{Tracking Node Wants (TNW)}%
\label{ssub:node_wants}

The TNW attack revolves around tracking which data items a given target node is interested in.
With the current implementation of the \ac{ipfs} node software,
nodes broadcast CID requests to \emph{all} of their connected peers (cf. \cref{sub:bitswap_explanation}, \cite{henningsen2020mapping}).
It is therefore sufficient for the adversary to maintain a connection to the target node and collect the requests it broadcasts.
From a practical standpoint,
a more challenging aspect of the TNW attack is to firstly determine the node ID of the target
This is also possible by again leveraging \ac{ipfs} itself:
the IDW attack can be used to discover nodes that are requesting some CID only the victim is likely to know or be interested in
.
In \cref{sub:gateways}, we demonstrate the effectiveness of this approach on well-known public gateway nodes on the \ac{ipfs} network.

\subsubsection{Testing for Past Interests (TPI)}%
\label{ssub:cache_testing}

In the TPI attack, the adversary seeks to confirm that a given node has recently accessed a given data item.
The attack leverages the fact that \ac{ipfs} nodes \emph{cache} previously requested data items locally and \emph{serve} them to interested parties~(cf. \cref{sub:content_retrieval}).
This caching mechanism is a cornerstone to the scalability and censorship-resistance of \ac{ipfs}.
However, in its current form it also enables any adversary capable of joining the \ac{ipfs} network to test whether a given target node has previously requested a given CID---by sending a request to the target node.
The target node will answer if the sought data item is in its cache,
and the data item will only be in the target node's cache if it was either requested or initially uploaded via the target node based on a user request.
Like for the TNW attack, the TPI attack can be mounted with negligible resources on part of the adversary, given the node ID of the target is known.

\subsection{Proof of Concept: Tracking Gateway Requests}%
\label{sub:gateways}

\ac{ipfs} offers a bridge to access the \ac{ipfs} network and hosted content through HTTP.
While every node offers this translation locally by default,
there are also a number of \emph{public gateways} available on the regular HTTP-based web,
maintained by the developer community and a number of organizations.
Public gateways are a convenient way to access \ac{ipfs} and aim to boost adoption of \ac{ipfs}, showcase examples, and enable access to the network in situations when running nodes locally is infeasible.
Public gateways also allow us to test privacy attacks in a realistic setting without threatening individual users.
In the following, we outline a methodology that combines the IDW attack with a tailored probing step for linking well-known public gateway nodes
(identified by their DNS names)
to \ac{ipfs} nodes (with node IDs and sometimes different IP addresses than the associated HTTP endpoints).
We apply the TNW attack on the identified nodes and briefly discuss the results of this investigation, which might be of independent interest.
The success of the presented experiment underlines the viability of the proposed attacks and the latent privacy risks they imply for \ac{ipfs} users.

\subsubsection{Gateway Probing Methodology}
\label{ssub:gateway_probing_setup}
We build upon our monitoring methodology (\cref{sec:monitoring}), extending it by a \emph{probing} step.
We generate a unique block of random data, yielding a unique \ac{cid} $c$\footnote{
  Obtaining a \ac{cid} duplicate is improbable due to the fact that \acp{cid} are based on cryptographic hashes of the data
  (s.a. \cref{sec:ipfs_in_a_nutshell}).
}.
We then launch a IDW attack on $c$,
adding our monitoring nodes $W$ as providers for $c$ to the \ac{ipfs} \ac{dht}.
Subsequently, we ourselves request $c$ through the HTTP-facing side of a public gateway and wait for \bs{} messages to arrive.
The received \bs{} request for $c$ enables us to uniquely identify the \ac{ipfs} side of the probed gateway.
Since $c$ refers to a block of randomly generated data, it is unlikely that any other user on the network would request $c$,
yielding a large confidence in the measured results.

\subsubsection{Gateway Probing Results}%
\label{ssub:gateway_probing_results}

We applied our gateway probing method to a list of public gateways curated by the \ac{ipfs} project\footnotemark.
We repeated the measurements two times, on May \nth{31} and June \nth{8} 2020 from two distinct hosts situated in Germany and the U.S., respectively.
From August 2021 we began searching for gateways regularly from the German monitor.
For each search and gateway we used a different random seed so that $c$ is unique in each trial.
The number of functioning gateways varies over time, as does the list of public gateways.
However, our results with regard to gateway functionality were always in line with the public list\footnotemark[\value{footnote}].
Using our probing method,
we were able to discover node IDs for all functional public gateways,
receiving relevant \bs{} messages from them at least once.
We also discovered node IDs for some of the non-functional gateways, i.e.,
our HTTP requests to them did not succeed but we still received relevant \bs{} messages.
We suspect a misconfiguration on the HTTP end of those gateways.

\footnotetext{
  \url{https://ipfs.github.io/public-gateway-checker/}
}

After collecting data through our probing step,
we cross-referenced the discovered IPs and overlay IDs with recent peer lists from our monitoring nodes $W$,
focusing on discovering nodes that share IP addresses and node IDs associated with multiple IP addresses.
The repeated probing and cross-referencing uncovered that several public gateways in above list were in fact associated with not just one but multiple nodes on the \ac{ipfs} network.
We contacted the operators of one prominent gateway with 13 associated \ac{ipfs} nodes, %
who confirmed that %
we correctly identified all of their nodes.
In total,
we discovered $93$ node IDs corresponding to well-known public gateways.
This number is growing over time as operators add new gateways to the public list, or regenerate their node IDs.
Because we perform gateway probing regularly based on the public list, we are able to track both changes to the list as well as changes to the individual gateways.

\subsubsection{Public Gateway Requests}
\label{ssub:gateway_usage_results}

\begin{figure}[ht]
	\centering%
	\input{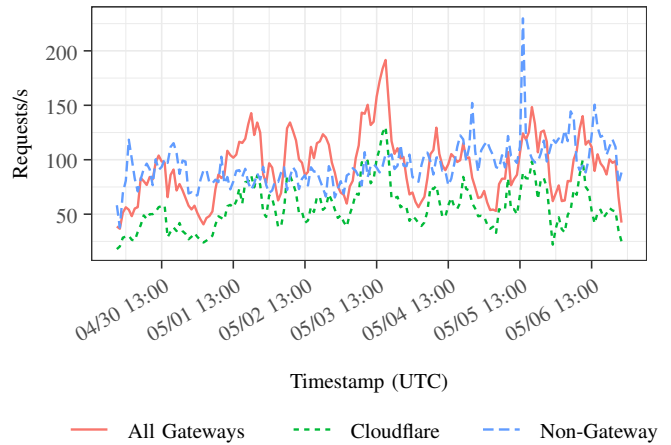}\unskip%
	\vspace{-0.2cm}%
	\caption[Deduplicated \bs{} Request Rate by Origin Group]{
		Deduplicated \bs{} request rate by origin group.
		The rate is calculated over one-hour slices of the unified, deduplicated trace (cf.\cref{sec:measurement_setup:data_processing}).
	}%
	\label{fig:wantlist_request_rate_no_dups}
\end{figure}

After successfully discovering the \ac{ipfs} nodes associated with well-known public gateways,
we can launch a TNW attack on these gateways.
In the following, we discuss results
collected \detailedweek{}.
\Cref{fig:wantlist_request_rate_no_dups} depicts the number of \bs{} requests per second that our two monitors received during this period.
We unified and deduplicated the traces from the two monitors as per \cref{sec:measurement_setup:data_processing}.
We plot both gateway and non-gateway (\enquote{homegrown}) request rates,
illustrating that we can successfully track the requests sent by a target node population.
We find that a significant portion of gateway traffic is due to a single public gateway operator---Cloudflare---and mark Cloudflare traffic explicitly in our results.

As can be seen, %
we in fact received a similar number of requests from both non-gateway nodes and gateway nodes.
This observation indicates that gateway \emph{usage} is a significant part of total \ac{ipfs} usage.
Furthermore, public gateways cache requested data at least as aggressively as regular nodes---if data requested via HTTP is already cached, no \bs{} request is generated and we thus cannot learn of this request.
Cloudflare states\footnote{In a keynote at the DI2F workshop at IFIP Networking 2021.} that their \ac{ipfs} gateway has a cache hit ratio of \SI{97}{\percent}, \ie, only \SI{3}{\percent} of requests
are forwarded immediately as \bs{} requests to the network%
\footnote{Back-of-the-envelope calculations based on the discussed data suggest that Cloudflare's gateways might be processing more than an order of magnitude more requests per second than the entirety of all monitored non-gateway nodes.}.
However, content is re-validated when its time-to-live expires which will potentially trigger a \bs{} request for the respective content.
This implies that our monitors are able to observe even cached \acp{cid} but are unable to infer their inter-request-time distribution.

\subsection{Privacy enhancement for decentralized data storage systems}%
\label{sub:privacy_countermeasures}

The feasibility and
relevance of the IDW, TNW, and TPI attacks are due to a number of inherent characteristics of \ac{ipfs}:

\begin{enumerate}
  \item\label{reason:ids} Long-lasting node identifiers---as connections must be maintained, nodes retain their node ID and IP address(es) for extended periods of time.
  \item\label{reason:connect_all} No connection limits---there are no mechanisms in place that can reliably prevent a small set of (adversary or monitoring) nodes from maintaining connections with nearly all nodes in the network,
    or a single adversary node from connecting to specific victim nodes.
  \item\label{reason:broadcast} CID request broadcasts---as part of the BitSwap protocol, nodes broadcast data requests to all nodes they are connected to.
  \item\label{reason:plain_cids} Plaintext CIDs---all recipients of data request learn the CIDs of requested data, not just nodes that can actually provide the data.
  \item\label{reason:cooperative} Cooperative caching---nodes cache downloaded data and cooperatively serve it to other interested nodes.
  \item\label{reason:no_cover} Single-user nodes with no cover traffic---users accessing the \ac{ipfs} network via a locally installed \ac{ipfs} client are represented in the network as a node that relays only their own and actual requests.
\end{enumerate}

The weakening of these attack enablers poses a significant challenge for designers of decentralized data storage systems.
Naive countermeasures can easily result in a significant deterioration of performance, censorship-resistance, or peer-to-peer scalability.
For example:

\begin{enumerate}
  \item Node identifiers can be cycled more frequently and an additional anonymization layer for IP addresses can be used.
    The effective cycling of node identifiers (i.e., existing connections need to be torn down) essentially increases churn.
    Using an established IP address anonymization service like Tor~\cite{dingledine2004tor} limits the performance of the data storage system to that of the anonymization service.
    It is an open question how to best integrate IP address anonymization functionality into \ac{ipfs} itself
    and what the performance impact of such a change would be.
    Parallels can be drawn to the design of privacy-centric systems such as Freenet~\cite{clarke2002protecting,roos2014measuring}.
  \item Systems like \ac{ipfs} thrive on their openness, on allowing anyone to join the network and provide a node.
    Attempts to limit the amounts of connections a single node can maintain are difficult to design
    and limited in effect as adversaries can easily split their connections between multiple Sybil~\cite{douceur2002sybil} nodes.
    Introducing per-connection \emph{costs},
    e.g., by requiring continuous Hashcash-like~\cite{back2002hashcash} proofs of work from peers,
    will likely also result in a decreased population of honest nodes while being of uncertain effect against determined adversaries with access to cloud computing resources.
  \item Nodes could request data items only from nodes found via DHT queries,
    rather than from their whole overlay neighborhood.
    However, this would reduce \ac{ipfs}'s robustness against censorship attacks~\cite{DBLP:journals/corr/abs-2011-00874} while being of limited effect as (Sybil) adversaries can also place themselves at key locations in the DHT.
    On the opposite end,
    extending \bs{} to support multi-hop requests or even flooding might not be sufficient to confuse dedicated adversaries either,
    as the identification of messages' sources is provably feasible even in decentralized flooding-based networks~\cite{neudecker2016timing}.
  \item The \bs{} protocol could be extended so that data requests contain only salted cryptographic hashes of CIDs together with the used salt value.
    Recipients of data requests would then need to calculate a salted hash for each CID they store in order to determine if they are a provider for the sought data.
    This approach would prevent adversaries from linking requests for data for which they do not know the CID.
    However, it would also induce additional computational overhead at providers
    and open up an effective amplification angle for denial-of-service attacks\footnote{
      The processing of data requests at providers would, without additional measures,
      induce a significantly higher workload at providers than at the nodes sending the requests.
    }.
    The described solution furthermore protects only the \bs{}-part of content retrieval in \ac{ipfs}
    (recall \cref{fig:content_retrieval_block_diagram}):
    plaintext CIDs would still need to be included in \ac{dht} queries in order for \ac{dht} routing to work correctly.
    Lastly, publicly-known CIDs, for example CIDs inferable from \texttt{ipfs://} URLs found on the Web,
    can still be tracked by adversaries even if CID hashing is used.
  \item Users can manually remove problematic items from their node's cache
    or configure their node to not reprovide downloaded items.
    While both measures are helpful against TPI attacks,
    they require manual user action and have no effect on IDW and TNW attacks.
    Disabling the reprovision of downloaded items
    would furthermore deteriorate censorship resistance of the network.
  \item Adding realistic cover traffic to add plausible deniability to one's genuine data requests is hard---in order for fake data requests to be effective they must be directed at actually existing CIDs and follow realistic popularity distributions.
    Lists of existing CIDs and their popularity distribution might be obtainable for monitoring operators
    (cf. \cref{sub:content_popularity}),
    but is not usually the case for regular users.
\end{enumerate}

Users could also use \ac{ipfs} via a public gateway (cf. \cref{sub:gateways}).
While this measure is both highly effective in terms of privacy-protection
and already in active use by a large part of the \ac{ipfs} user base
(cf. \cref{ssub:gateway_usage_results})
it arguably weakens the benefits of \ac{ipfs} as a \emph{decentralized} data storage system.
Namely, by accessing content on \ac{ipfs} without participating in the \ac{ipfs} network,
users do not contribute to the network's scalability and censorship resistance.

Turning the gateway strategy on its head,
regular \ac{ipfs} users could also \emph{provide} gateway services,
thereby both strengthening the \ac{ipfs} network and adding natural cover traffic to requests they themselves send via their node.
For increasing the use of smaller gateway services,
changes to the \ac{ipfs} software are necessary so that the selection of a gateway is handled automatically. %
The solution can be expanded towards a form of onion routing~\cite{dingledine2004tor},
with data requests being routed through several gateways and only the last gateway in the chain performing an actual DHT- and BitSwap-based search,
or other deniability-increasing approaches like AP3~\cite{mislove2004ap3}.

We leave the in-depth investigation of this and other potentially promising privacy-hardening approaches for future works.
On a side note, providing gateway services to other (anonymous) users leads to some legal uncertainties.
Gateway nodes can be led into caching data the mere possession of which might be considered a criminal act.
Depending on jurisdiction, providers of hosting services are often exempt from direct liability as long as they quickly remove problematic data upon obtaining knowledge or awareness of its existence
(see for example \cite[Art. 14]{eu2000ecommerce})
and it is plausible that such safe harbor rulings would extend to small-scale \ac{ipfs} gateways.

\section{Conclusion}%
\label{sec:conclusion}

We presented a novel monitoring methodology for observing data-related activity in \ac{ipfs},
a highly popular decentralized data storage system.
Leveraging inherent features of \ac{ipfs}' design, we are able to maintain connections with large portions of the node population at low cost,
and receive indicators for all \acp{cid} that connected-to nodes request and look for.
We tested our methodology in a fifteen month measurement study and discuss exemplary results about the \ac{ipfs} network.
For a studied 7-day period,
roughly \SI{70}{\percent} of activity can be attributed to just three countries and the popularity of content items does not follow a power-law distribution.
Despite its potential for generating informative insights about the \ac{ipfs} network,
the effectiveness of our monitoring approach also paints a troubling picture with regard to user privacy.
We describe how our methodology can be extended for realizing privacy attacks that ultimately enable low-resource adversaries to surveil the current and past data requesting behavior of node-operating users.
While we demonstrate the practical viability of our attacks,
we also map directions for hardening \ac{ipfs} and similar systems without
significantly impacting other core features.
Last but not least, the collected data opens up avenues for further research, \eg, in-depth investigations into the filesystem layer of \ac{ipfs} and the content stored on the network.

\section{Acknowledgments}
\label{sec:acknowledgements}
This work was funded in part by grant PL-RGP1-2021-054.

\printbibliography

\end{document}